\documentclass[]{elsart}
\usepackage{amsmath}
\usepackage{amssymb}
\usepackage{graphicx}
\usepackage{xspace}               %% Provides a conditional space after commands
\usepackage{subfigure}            %% Subfigures

\newcommand{\mevcc}{\ensuremath{\mathrm{MeV}/c^2}\xspace}
\newcommand{\mevc}{\ensuremath{\mathrm{MeV}/c}\xspace}

\newcommand{\gevcc}{\ensuremath{\mathrm{GeV}/c^2}\xspace}
\newcommand{\gevc}{\ensuremath{\mathrm{GeV}/c}\xspace}

\newcommand{\mathsl}[1]{\mbox{\textsl{#1}}}
\newcommand{\meson}[1]{\ensuremath{\mathsl{#1}}}
\newcommand{\kaon}{\ensuremath{\meson{K}}\xspace}
\newcommand{\kshort}{\ensuremath{\kaon_{s}^{0}}\xspace}
\newcommand{\xim}{\ensuremath{\Xi^-}\xspace}

\graphicspath{{figs/} }

\begin{document}

\begin{frontmatter}

\title{Search for a pentaquark decaying to $\Xi^- \pi^-$}

\input{author_list.plb}%%

\begin{abstract}
We present a search for a pentaquark decaying strongly to $\Xi^-\pi^-$ in $\gamma N$ 
collisions at a center-of-mass energy up to 25~GeV/$c^2$.  Finding no
evidence for such a state in the mass range of 1480~MeV/$c^2$ to 2400~MeV/$c^2$, 
we set limits on the yield and on the cross section 
times branching ratio
relative to $\Xi^*(1530)^0$.
\end{abstract}


\end{frontmatter}

\section{Introduction}

The existence of bound multiquark states like $Q\overline{Q}q\overline{q}$ and
the $H$ dihyperon were first proposed by Jaffe 
\cite{Jaffe:1976ig,Jaffe:1976ih,Jaffe:1976yi}
in 1977. 
Then 20 years later Diakonov \textit{et al.} 
\cite{Diakonov:1997mm}
proposed the 
existence of 
four quarks and one antiquark confined in a low-mass anti-decuplet 
configuration.  In their calculations
Diakonov made several predictions of masses and widths of exotic baryonic states 
such as the mass of the lightest state $\Theta^+$ (previously called $Z^+$) 
at about 1530 \mevcc.
There were also predictions of decay modes: $\Theta^+ \to p \kshort$, 
$\Theta^+ \to n K^+$ and
$\Xi_5^{--} \to \xim \pi^-$
\footnote{The $\Xi_5^{--}$ is also known as the $\phi(1860)^{--}$.}.
 
On the experimental side, the year 2003 was 
the beginning of ``pentaquark observations.'' The $\Theta^+$ at about the
predicted mass was the first candidate as noted in the PDG2004
\cite{Eidelman:2004wy}, but searches  by higher statistics experiments
yielded negative results as noted by 
R. Schumacher \cite{Schumacher:2005wu}.
 
In the case of the doubly strange pentaquark $\Xi_5^{--}$ the only
positive evidence is from NA49 
\cite{Alt:2003vb}.  Negative results were obtained by every other search: 
HERA-B \cite{Abt:2004tz}, ALEPH \cite{Schael:2004nm}, WA89 \cite{Adamovich:2004yk},
HERMES \cite{Airapetian:2004mi}, BABAR \cite{Aubert:2005qi}, ZEUS \cite{Chekanov:2005at}, 
COMPASS \cite{Ageev:2005ud}, 
E690 \cite{Christian:2005kh}, as well as preliminary results from 
CDF \cite{Litvintsev:2004yw}.
                                                                                
This letter describes a search for the $\Xi_5^{--} \to \xim \pi^-$
pentaquark candidate\footnote{Charged conjugate states are implied unless explicitly
stated otherwise.} in electromagnetic interactions and extends the search
for the singly strange $\Theta^+$ discussed in \cite{Link:2006yh} to the doubly strange
state considered here.

\section{Event reconstruction and selection}

\begin{figure}[h]
\centerline{
\includegraphics[width=4.7in,height=2.3in]{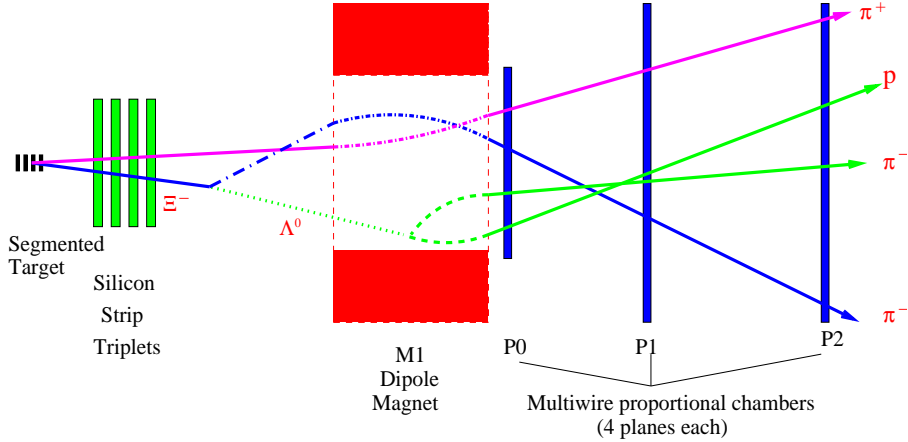}
}
\caption{
A schematic drawing in the bend view of the spectrometer of a 
 $\Xi^0(1530) \to \Xi^-\pi^+$ decay, with the $\Xi^-$ decaying downstream 
the silicon strip detector, as described in the text. Only the front part of the
spectrometer is displayed.
}
\label{fig:schematic}  % fig 1 new
\end{figure}

The FOCUS experiment took data during the 1996--7 fixed-target run at Fermilab.  
A photon beam obtained from bremsstrahlung of 300~GeV electrons and positrons impinged
on a set of BeO targets.  
%Four sets of silicon strip detectors, each with three views, 
%were located downstream of the targets for vertexing and track finding.  
The first element in the  spectrometer was the silicon strip detector array: four triplets of
silicon strip planes used for track finding and vertexing.  
Each triplet was comprised of three closely spaced parallel planes with the
strip directions rotated to provide a means for correlating hits in the
three planes of the triplet and thus track coordinates.
For most of 
the run, two pairs of silicon strips were also interleaved with the target segments for more
precise vertexing~\cite{Link:2002zg}.  Charged particles were tracked and momentum 
analyzed as they passed
through one or two dipole magnets and three to five sets of multiwire proportional chambers 
with four planes each (as shown in Fig. \ref{fig:schematic}).  
Three multicell threshold \v{C}erenkov counters, two electromagnetic
calorimeters, and two muon detectors provided particle identification.  A trigger which 
required, among other things, $\gtrsim$25~GeV of hadronic energy passed 6 billion events for 
reconstruction.

\begin{figure}[h]
\centerline{
\includegraphics[width=2.7in,height=2.3in]{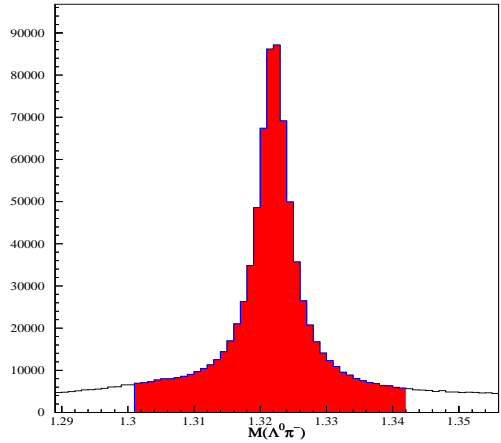}
}
%\hspace{-3pt}
\caption{
The invariant mass plot represent the total sample of silicon tracks matched to
a reconstructed $\Lambda^0$ and a multiwire chamber track ($\pi-$).
The shadow area is the region used to select \xim tracks for this analysis.   
}
\label{fig:cas}  % fig 2 % old fig 1
\end{figure}

The data used for this analysis come from a subset of FOCUS data which contain cascade  
candidates 
($\Xi^- \to \Lambda^0 \pi^-$ and $\Omega^- \to \Lambda^0 K^-$).  
The cascade 
decays used in this analysis are those which occur downstream of the silicon 
detector and with the $\Lambda^0$ daughter being fully reconstructed through the decay 
$\Lambda^0 \to p\pi^-$.
The $\Xi^-$ candidate is a reconstructed silicon track with direction and position consistent with the 
intersection of a reconstructed $\Lambda^0$ and a multiwire chamber track.
The $\Lambda^0$ invariant mass is 
required to be between 1.10 \gevcc and 1.13 \gevcc. The higher momentum track is chosen 
to be the proton because the small phase space of the $\Lambda^0\to p \pi^-$ 
constrains the proton to carry most of the momentum for the decays observed
in the forward FOCUS spectrometer.  
For this analysis we identify $\Xi^-$ tracks when the matched
$\Lambda^0$ candidate and the matched pion track have an invariant mass 
within 20 \mevcc of the nominal $\Xi^-$ signal peak (shadow region of fig.~\ref{fig:cas}).
The total sample yields approximately  625\,000 $\xim \to \Lambda^0\pi^-$ 
signal events. %as shown in Fig.~\ref{fig:cas}.
A detailed description of the reconstruction of cascades and vees 
can be found in Ref.~\cite{Link:2001dj}.  

Each $\Xi^-$ track is combined with good quality charged tracks to make a vertex
to search for $\Xi^*(1530)^0 \to \xim \pi^+$ and $\Xi_5^{--} \to \xim \pi^-$.
This two track vertex must be well defined with a $\chi^2$\ probability greater than
1\%. A multitrack production vertex is nucleated around this two track vertex 
and must be within $2\sigma$ of the $\xim\pi$ vertex.
The production vertex
must have a $\chi^2$\ probability greater than 1\% and must be within
the target material or outside by no more than $3\sigma$. 
In both cases, $\sigma$ is the calculated uncertainty on the vertex location or separation.

A particle identification
algorithm has been developed which combines data from all of the
\v{C}erenkov counters which the track passes through.   This algorithm~\cite{Link:2001pg} 
returns negative log-likelihood  (times two) values $W_i$ for a track and 
hypothesis $i\in\left\{e,\pi,K,p\right\}$ 
based on the light yields in the phototubes covering the \v{C}erenkov cone of the track.
A \v{C}erenkov cut,
$W_{\textrm{min}(e,K,p)}-W_{\pi} > -8$, requires that 
the pion \xim daughter ($\pi^-$ track matched to \xim track in Fig.~\ref{fig:schematic})
must not be strongly inconsistent with the pion 
hypothesis.  
For pions of the $\xim\pi$ combination ($\pi^+$ track in Fig.~\ref{fig:schematic})
a \v{C}erenkov cut, $W_{\textrm{min}(e,K,p)}-W_{\pi}>-6$, is used to
reduce combinatorial background.

\begin{figure}[h]   % fig 3 % old fig 2
\centerline{\includegraphics[width=5.5in]{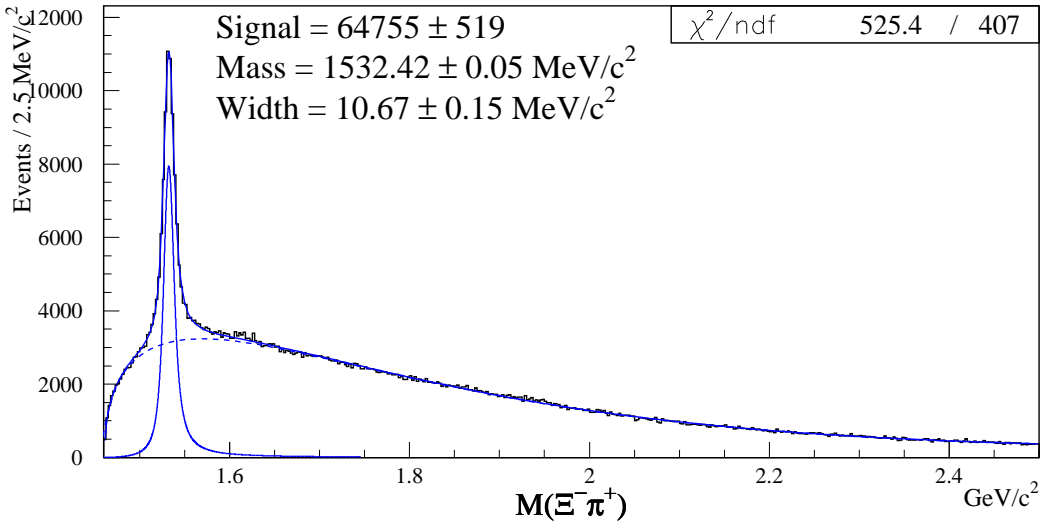}}
\caption[$\Xi^*(1530)^0$ fit with P-wave Breit-Wigner]{$\Xi^*(1530)^0$ fit with
a P-wave Breit-Wigner and 
combinatorial background.}
\label{fig:casstar}
\end{figure}

\begin{figure}[h]   % fig 4  % old fig 3
\centerline{\includegraphics[width=5.5in]{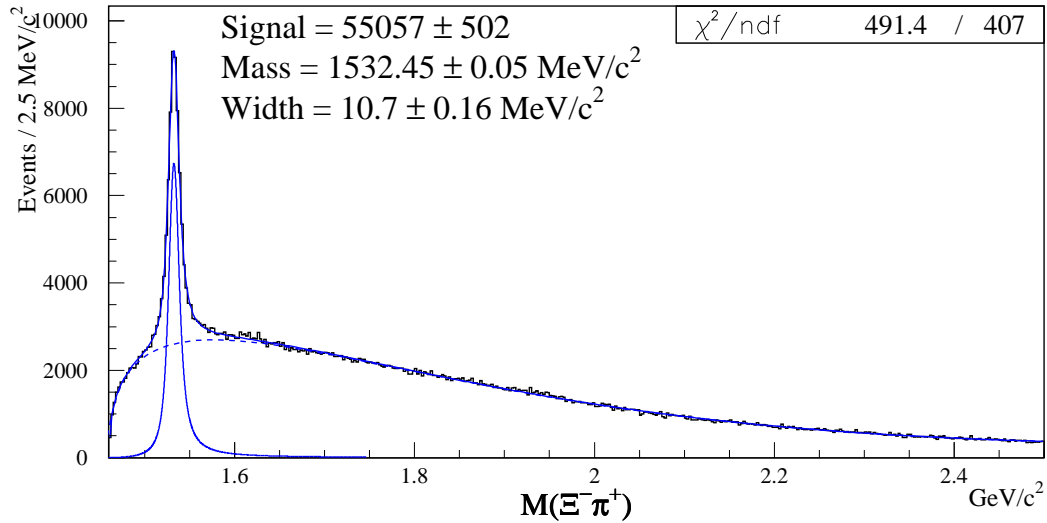}}
\caption[$\Xi^*(1530)^0$ fit with P-wave Breit-Wigner for $p>$25\gevc]{$\Xi^*(1530)^0$ fit with a P-wave 
Breit-Wigner for momentum greater than 25 \gevc and
combinatorial background.}
\label{fig:casstarp25}
\end{figure}

Mass plots for $\xim\pi^+$
 are shown in Fig.~\ref{fig:casstar} and in Fig.~\ref{fig:casstarp25} with an additional cut that
the $\xim\pi^+$ momentum be greater than 25 \gevc.
The signal was best fit with a P-wave Breit-Wigner with an energy dependent width 
convoluted with a Gaussian for the detector resolution.  The resolution,
2.92 \mevcc, was obtained from a Monte Carlo simulation and the fitted width of
the Breit-Wigner is shown in the figures, consistent with the widths of 8-10 \mevcc
quoted in the PDG\cite{Eidelman:2004wy}.  
The background was fit to the form $a q^b \exp{(cq + dq^2 + eq^3 + fq^4)}$ where 
$a$---$f$ are free parameters and $q$ is the $Q$-value (invariant mass minus
component masses).  

\section{Pentaquark search results}

\begin{figure}  %fig 5 %%old fig 4
\centerline{\includegraphics[width=5.0in]{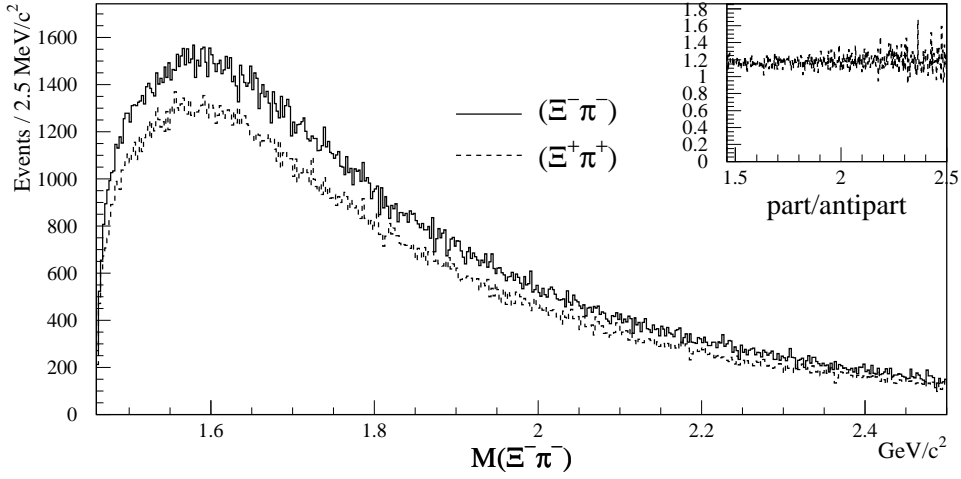}}
\caption{
The invariant mass distribution of $\xim \pi^-$ separated by charge (particle and antiparticle).
   Standard cuts are applied (no momentum cut). The inset plot is the ratio of particle/antiparticle.
}
\label{fig:result:wrongsign2}
\end{figure}

\begin{figure}  %fig 6 %%old fig 5
\centerline{\includegraphics[width=5.0in]{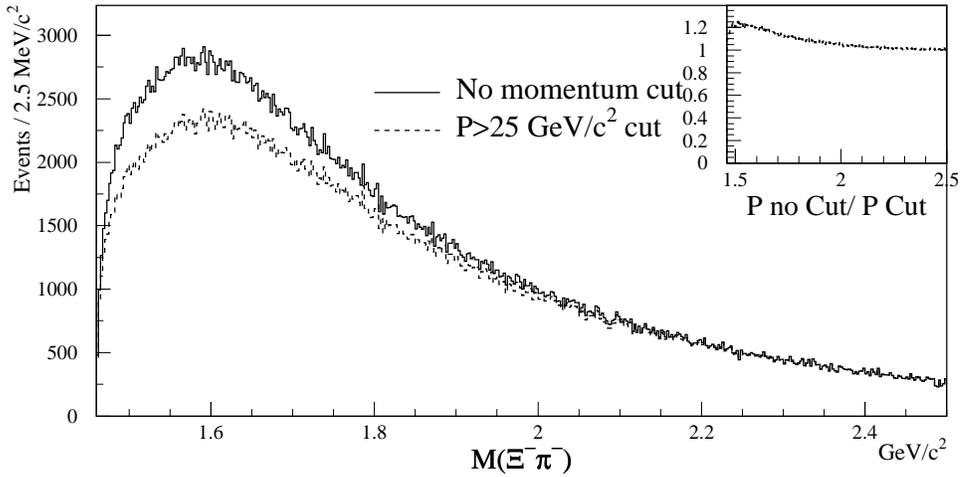}}
\caption{The invariant mass distribution of $\xim \pi^-$ for particle and 
antiparticle combined. The solid line
shows the result for the standard cuts and the dashed line is with the 
additional cut that the momentum is greater than 25 \gevc. The inset plot is the ratio of
both, note that the momentum cut affects primarily low mass.
}
\label{fig:result:wrongsign1}
\end{figure}

\begin{figure}  % fig 7 %%old fig 6
\centerline{\includegraphics[width=2.75in]{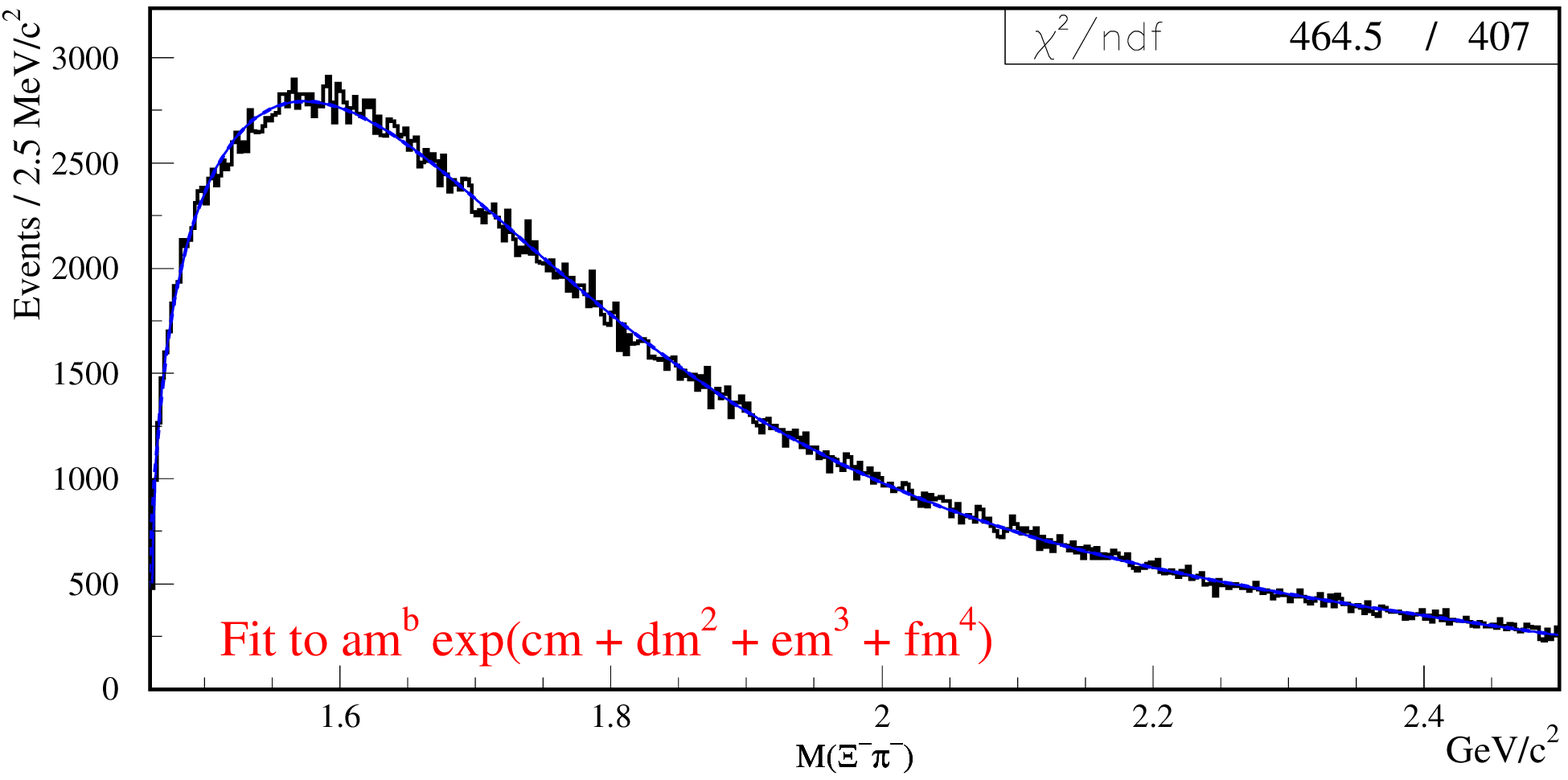}
\includegraphics[width=2.75in]{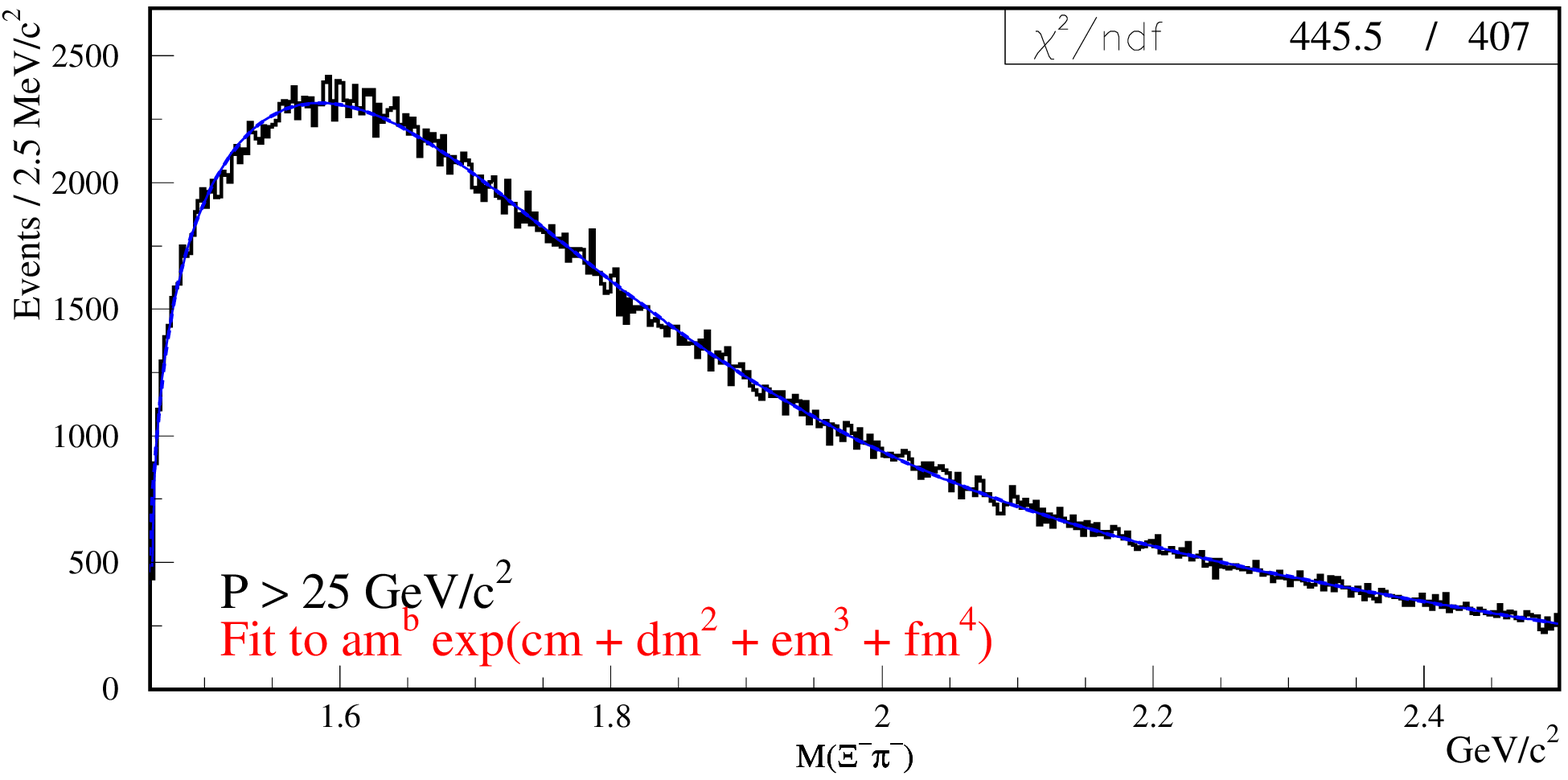}}
\caption{Fit to the invariant mass distribution of $\xim \pi^-$ for particle and antiparticle 
combined. The left figure
show the result for the standard cuts and the right figure 
with the additional cut that the momentum is greater than 25 \gevc.
}
\label{fig:result:wrongsign_fit}
\end{figure}

The $\Xi^-\pi^-$ and $\Xi^+\pi^+$ invariant masses are plotted using the standard selection 
criteria in 
Fig.~\ref{fig:result:wrongsign2}.  
There are no significant differences between the two charge states so for 
the remainder of the analysis we combine the charge conjugate states.
The combined sample with standard cuts and with an additional momentum
cut of 25 \gevc is plotted in Fig.~\ref{fig:result:wrongsign1}.  
In this analysis we treat the two samples displayed in
Fig.~\ref{fig:result:wrongsign1}  separately, 
because the production mechanism utilized in Monte Carlo
acceptance calculations is much better understood at the higher momenta.
The same parameterization of the
background is used in Fig.~\ref{fig:result:wrongsign1} as in Fig.~\ref{fig:casstar} and \ref{fig:casstarp25}.
%
%In both cases of Fig.~\ref{fig:result:wrongsign1} the total sample is 
%fit to a background curve of the form 
%$a q^b \exp{(cq + dq^2 + eq^3 + fq^4)}$ where $a$---$f$ are free parameters and 
%$q$ is defined as $q \equiv M(\Xi \pi)-m_{\Xi}-m_{\pi}$. 
%
%The particle and antiparticle mass spectra are combined in
%Fig.~\ref{fig:result:wrongsign_fit} and show acceptable
%fits to this smooth background.
Such parameterization show acceptable fits to this smooth background as shown in Fig.~\ref{fig:result:wrongsign_fit}.
No evidence for a pentaquark near $1860$~MeV/$c^2$ or at any mass less than $2400$~MeV/$c^2$ is observed.
To set a limit on the yield we need to make some assumptions about the width of the state.  
We consider two cases:
one with a natural width of 0 and one with a natural width of 15 \mevcc.\footnote{
NA49 detected this state with a width below the  detector resolution of 18 \mevcc. Therefore
    the expected width of this state cannot exceed this value. A choice of natural width of 15 \mevcc
convoluted with FOCUS resolution in a range of 4-11 \mevcc gives about 18 \mevcc, the upper limit
set by NA49.
}  
In the first case, the signal is
fit with a Gaussian with a width given by the experimental resolution.  
In the second case, the signal is 
fit with a P-wave Breit-Wigner with an energy dependent width convoluted with the experimental
resolution.  
The experimental resolution $\sigma$ is parametrized as a function of the invariant mass and 
we find $\sigma = -9.35 + 7.76m + 0.21m^2$ an adequate approximation, with $\sigma$ in \mevcc and $m$ is the mass in 
\gevcc.
With the additional momentum cut of 25 \gevc the experimental resolution change to 
$\sigma = -7.99 + 6.29m + 0.62m^2$.


\begin{figure}         %fig 7
\centerline{\includegraphics[width=5.5in]{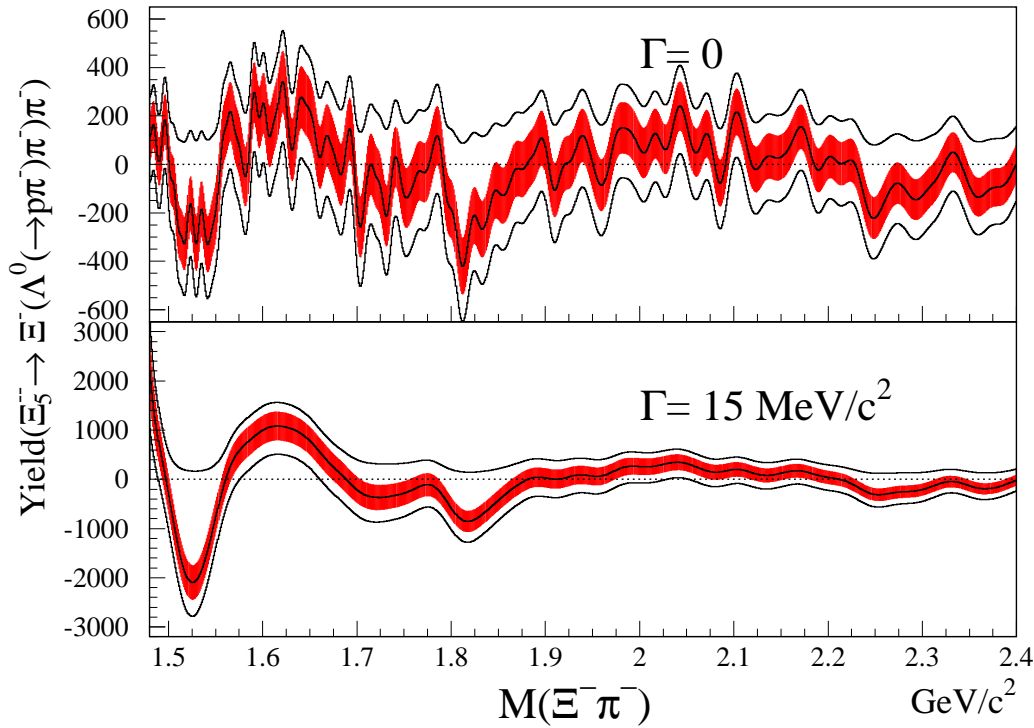}}
\caption{Pentaquark yields and upper limits.  Top (bottom) plots show
results for a natural width of $0$ ($15$~MeV/$c^2$).  The shaded region includes
the $1 \sigma$ errors with the central value in the middle.
The outer curves show the upper and lower 95\% confidence limits.}
\label{fig:result:penta_yld}
\end{figure}

\begin{figure}[htbp]    % fig 8
  \centerline{
  \includegraphics[width=5.5in]{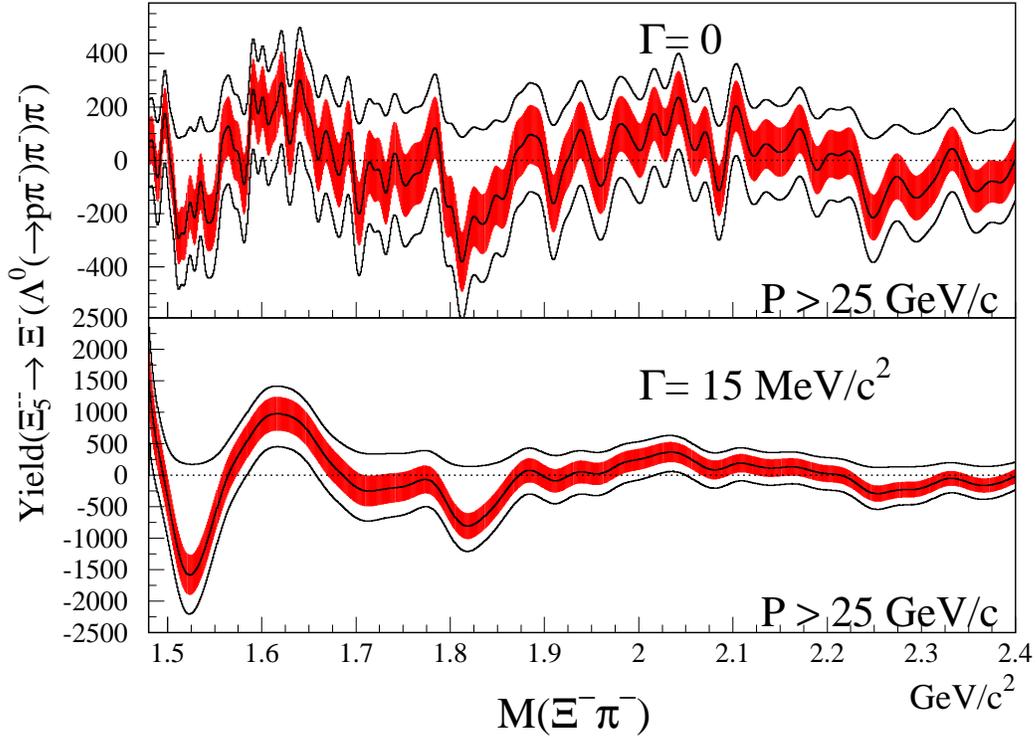}}
  \caption{
   Pentaquark yield and limits for standard cuts including the 25 \gevc momentum
 cut. Top (bottom) plots show
results for natural width of 0
   (15 \mevcc). The shaded region includes the $1 \sigma$ uncertainty with the 
central value in the middle.
   The outer curves show the upper and lower 95\% confidence limits.}
  \label{fig:result:penta_yldp25}
\end{figure}

A series of 921 fits to the observed $\Xi^-\pi^-$ mass plot were performed using the background and 
signal shapes described above for each assumed width.  
The signal mass is varied in 1 \mevc steps from $1480$ to $2400$ \mevcc and
a binned log-likelihood fit using \textsc{Minuit}~\cite{minuit} is performed.
The $\pm 1 \sigma$ errors are defined as the point where $\Delta \log{\mathcal{L}} = 0.50$ relative to 
the maximum $\log{\mathcal{L}}$, while continually adjusting the background parameters to maximize 
$\log{\mathcal{L}}$.  
The 95\% CL 
lower limit is defined similarly with $\Delta \log{\mathcal{L}} = 1.92$. 
Both are obtained using \textsc{Minos}~\cite{minuit}.  The 95\% CL 
upper limit is constructed as follows:  
The likelihood function $\mathcal{L}$ versus yield is determined by maximizing $\log{\mathcal{L}}$ for 
many different (fixed) yields, allowing background parameters to float.  The likelihood
function is integrated from a yield of $0$ to $\infty$ to obtain the total likelihood.  The 95\% CL 
upper limit on the yield is defined as the point where 95\% 
of the total likelihood is between a yield of $0$ and the upper limit.\footnote{  
This definition of an upper limit is used rather than a counting based
Feldman--Cousins type limit because errors are Gaussian for this large background.}
The fitted yield, $1 \sigma$ errors, and 95\% CL 
limits are shown in Figs.~\ref{fig:result:penta_yld} and \ref{fig:result:penta_yldp25}.  
Of the 1842 fits, none of them finds a positive excursion
greater than 5$\sigma$.  
The only previous pentaquark observation was around 1860~MeV/$c^2$.
In this region, we find a small dip which is not statistically significant.
The largest positive excursion occurs in the region where the background distribution peaks.


To compare with other experiments, the limits on yield must be converted to limits on production times the
(unknown) branching ratio.  We choose to normalize the $\Xi^{--}_5$ production 
cross section to $\Xi^*(1530)^0$ because the reconstructed decay mode of the
$\Xi^*(1530)^0 \to \Xi^- \pi^+$ is very similar,
in terms of topology and energy release, to the state we are investigating.  
Thus, we attempt to determine
\begin{equation}
 \frac{\sigma(\Xi_5^{--})\cdot BR(\Xi_5^{--} \to \xim \pi^-)}{\sigma(\Xi^*(1530)^0)} =
 \frac{Y(\Xi_5^{--})\cdot BR(\Xi_5^{--} \to \xim \pi^-)}{Y(\Xi^*(1530)^0)}\cdot
 \frac{\epsilon_{\Xi^*(1530)^0}}{\epsilon_{\Xi_5^{--} \to \xim \pi^-}}
\label{eq:relx}
\end{equation}

All of the efficiencies include the reconstruction and selection efficiencies and corrections for 
unseen decays of parent particles.  
The $\Xi^*(1530)^0$ efficiency includes 
BR$(\Xi^*(1530)^0\to \xim \pi^+)=0.66$ and both efficiencies include the factors
BR$(\xim \to \Lambda^0 \pi^- )=1$ and BR$(\Lambda \to p \pi^- )= 0.64$.
The two branching ratios which are common to both cancel.
Determining reconstruction
and selection efficiency (including acceptance) is described below.

The FOCUS detector is a forward spectrometer and therefore acceptance depends on
the momentum of the produced particle.  The production characteristics of the pentaquark are the
largest sources of systematic uncertainty in this analysis.  
We choose a particular production model to obtain limits and provide
sufficient information about the experiment for other interested parties to obtain limits based on other production 
models.  The production simulation begins with a library of $e^-$ and $e^+$ tracks obtained from a 
TURTLE simulation~\cite{turtle} of the Wideband beam line.  
From this library, an individual track is drawn and bremsstrahlung photons are created by passage through 
a 20\% $X_0$ lead radiator.  Photons with energy above 15~GeV are passed to the \textsc{Pythia}~\cite{Sjostrand:1993yb} 
Monte Carlo simulation. The \textsc{Pythia} version we use is 6.127.
The \textsc{Pythia} simulation is run using minimum bias events\footnote{
Specifically \texttt{MSEL=2}
in the \textsc{Pythia} setup.}
with varying energies.\footnote{
Specifically 
\texttt{MSTP(171)=1}
in the \textsc{Pythia} setup.} 
Options controlling parton distributions and gluon fragmentation were set to avoid heavy quark 
production.\footnote{ 
\texttt{MSTP(58)=3} to produce light quarks (uds)  only and 
\texttt{MDME(156--160,1)=0} to limit gluon fragmentation into light quarks only 
in the \textsc{Pythia} setup.
}
Since \textsc{Pythia} does not produce pentaquarks, another particle must be
chosen to represent the pentaquark.  According to the string fragmentation model,
which is implemented in \textsc{Pythia}, the mass of the particle has the greatest
effect on production and the number of quarks a particle has in common with the
initially interacting hadrons is next in importance.
The $\Xi^*(1530)^0$ particle is chosen to represent the production of a pentaquark.
The $\Xi^{*0}(ssu)$ can obtain at most 33\% 
of the remaining quarks from the target nucleon valence quarks, 
while the $\Xi_5^{--}(ddss\overline{u})$ can take 40\%.  
The charge conjugate $\overline{\Xi^*(1530)^0}(\overline{ssu})$
particles must obtain all quarks from the vacuum, while the
$\overline{\Xi^{--}_5}(\overline{ddss}u)$ can take 20\% from the target nucleon.
The mass of the particle chosen to represent the
pentaquark, $\Xi^*(1530)^0$, 
is set to the appropriate value in \textsc{Pythia}.\footnote{ 
By setting \texttt{PMAS(190,1)} in \textsc{Pythia}.
}

\begin{figure} % fig 9
\centerline{\includegraphics[width=5.5in]{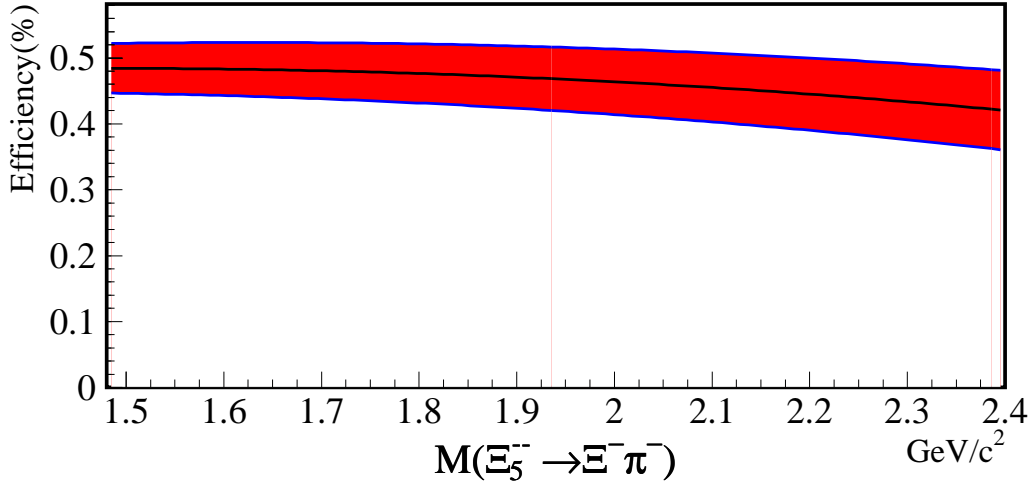}}
\caption{Acceptance versus mass for pentaquark candidates.  
Upper (lower) curve is for a 1$\sigma$ pentaquark produced uncertainty.
}
\label{fig:accvsm}
\end{figure}

To calculate the relative cross sections in Eq.~\ref{eq:relx} we need efficiencies for 
{$\Xi^*(1530)^0 \to \Xi^-\pi^+$} and {$\Xi_5^{--} \to \Xi^- \pi^-$}.  
These efficiencies are obtained from the FOCUS Monte Carlo simulation.  The 
dominant uncertainty in the efficiency determination is the modeling of the production 
characteristics of the parent particle.  For the observed particle, 
$\Xi^*(1530)^0$, we can compare the data and Monte Carlo directly
and adjust the Monte Carlo simulation to produce the correct data distribution.
Even this is 
not sufficient, however, because areas where the efficiency is zero cannot be accounted for.
For $\Xi^*(1530)^0$, we run a weighted Monte Carlo simulation which 
matches the Monte Carlo momentum distribution with the observed data momentum distribution in
the region for which the acceptance is not zero.  The dominant source of uncertainty for the
$ \Xi^*(1530)^0\to \Xi^- \pi^+$ efficiency is
our lack of knowledge of the fraction of events completely outside of our acceptance (momentum less
than 15~GeV/$c$).  The weighted \textsc{Pythia} Monte Carlo predicts that 71\% of the
$\Xi^*(1530)^0$ particles are produced with momentum less than 15~GeV/$c$.
To obtain an estimate of the efficiency uncertainty, we assume that the number of 
particles with momentum less than 15~GeV/$c$ can be off by as much
as a factor of 2 (high or low).  
This leads to a relative
uncertainty on the $\Xi^*(1530)^0$ efficiency of 45\%.
The $\Xi_5^{--}\to \Xi^-\pi^-$ efficiency is taken to match the uncertainty of a higher 
statistical mode $\Theta^+\to pK_S^0$ ($\sim$5\%) \cite{Link:2006yh}  when using a substitute particle like 
$\Xi^*(1530)^0$ and $\Sigma^*(1385)^+$ to generate pentaquarks with a momenta greater than 25 \gevc.
The $\Xi_5^{--}\to \Xi^-\pi^-$ efficiency versus mass (with no branching ratio corrections) 
is shown in Fig.~\ref{fig:accvsm}.
The uncertainty in $\epsilon_{\Xi_5^{--}\to \Xi^-\pi^-}$ is approximately 10\%.
The relative uncertainty of the efficiency of an unknown particle ($\sim$10\%) is indeed less
than that for the high statistics normalizing modes ($>$40\%) because the efficiency
uncertainty of the high 
statistics modes reflects the lack of knowledge of production outside of our acceptance.
It is reasonable to assume that discrepancies in the Monte Carlo simulation are similar for
the signal mode and the normalizing mode since discrepancies are correlated
and therefore adding the uncertainty to the signal mode
would be double-counting.  Note that the signal and normalizing efficiencies only appear as a ratio.

\begin{figure} %fig 10
\centerline{\includegraphics[width=5.5in]{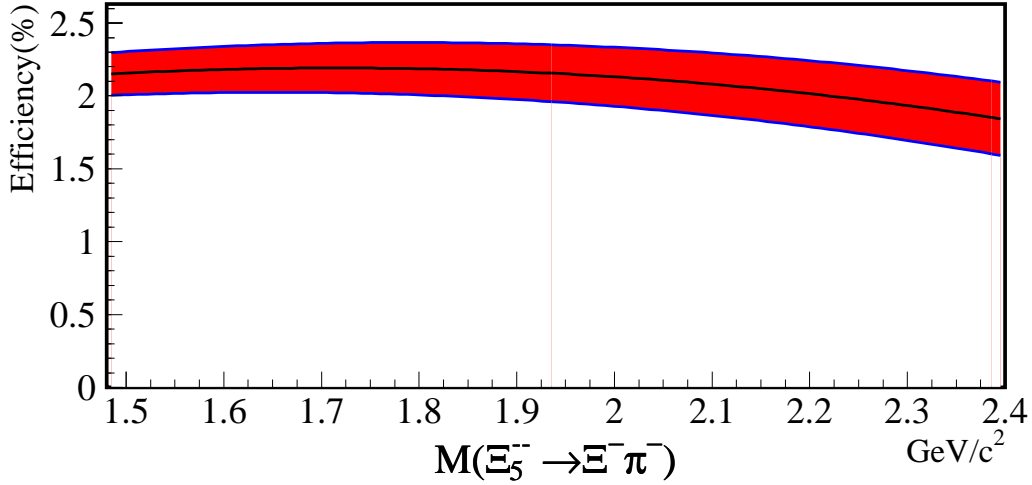}}
\caption{Acceptance versus mass for pentaquark candidates.  
Lower (upper) curve is for $1 \sigma$ pentaquark produced uncertainty.
The pentaquark is produced and reconstructed with momentum greater than 25~GeV/$c$.}
\label{fig:accvsma}
\end{figure}

We also report the relative cross sections in the region where our acceptance is good, that is for
parent particle momenta greater than 25~GeV/$c$.  This dramatically reduces the systematic uncertainties
associated with the measurement.  The uncertainty due to 
the production of $\Xi^*(1530)^0$ is minimal.  The uncertainty in the
$\Xi_5^{--}$ efficiency is also reduced from approximately 10\% to about 5\% as shown in 
Fig.~\ref{fig:accvsma}.  The number of reconstructed  
$\Xi^*(1530)^0$ at momenta greater than 25~GeV/$c$ is about 55\,000, 
compared to a total sample of about 65\,000 without the momentum cut.

The upper limit on the yield was obtained by mathematically integrating the likelihood function from 0 to 
infinity and then integrating from 0 to 95\% of the total likelihood 
integral to obtain the 95\% CL upper limit.
To obtain the limit on cross section requires a different approach because of the significant systematic 
uncertainties.  We use a method based on a note by Convery~\cite{Convery:2003af} which is inspired by the
Cousins and 
Highland~\cite{Cousins:1991qz} philosophy for including systematic uncertainties.  
The Cousin and Highland 
prescription is appropriate for low background experiments with Poisson errors while the Convery proposal
is applicable to the Gaussian errors which result from the large background in our case.
Modifications to the Convery approach are made to give an exact solution~\cite{stenson} when we include 
efficiency uncertainty systematics.  

If systematics are not considered,
an analysis using a maximum likelihood fit returns a central value for the branching ratio $(\hat{B})$
and a statistical error $(\sigma_B)$.  The likelihood function is
\begin{equation}
p(B) \propto \exp{\!\left[\frac{-(B-\hat{B})^2}{2 \sigma_B^2}\right]}.
\end{equation}
Including the uncertainty on the efficiency ($\sigma_\epsilon$) changes the likelihood to:
\begin{multline}
\label{eq:fulleps}
p(B) \propto \frac{1}{\sqrt{\frac{B^2}{\sigma_B^2}+\frac{1}{\sigma_\epsilon^2}}}
\exp{\!\left[\frac{-(B-\hat{B})^2}{2(B^2\sigma_\epsilon^2+\sigma_B^2)}\right]}\left\{
\textrm{erf}\!\left[\frac{B \hat{B} \sigma_\epsilon^2 + \sigma_B^2}{\sqrt{2}\sigma_\epsilon \sigma_B 
\sqrt{B^2\sigma_\epsilon^2 + \sigma_B^2}}\right]\right. - \\
\left. \textrm{erf}\!\left[\frac{(\hat{S}-1)\sigma_B^2 - B\sigma_\epsilon^2 (B - 
\hat{B}\hat{S})}{\sqrt{2}\hat{S}\sigma_\epsilon \sigma_B \sqrt{B^2 \sigma_\epsilon^2 + \sigma_B^2}}\right]\right\}.
\end{multline}

We integrate Eq.~\ref{eq:fulleps} from 0 to $\infty$ to obtain the total probability and then integrate from 0 
to the point at which 95\% of the total probability is included to obtain the 95\% CL upper limit.
The branching ratio $B$ of Eq.~\ref{eq:fulleps} is simply the relative cross section times the unknown 
pentaquark branching ratio as in Eq.~\ref{eq:relx}.  The relative uncertainties on the efficiency for the 
signal and normalizing mode are added in quadrature to become $\sigma_\epsilon$ in Eq.~\ref{eq:fulleps}.
Furthermore, $\hat{S}$ is the relative efficiency between the signal and normalizing modes and $\sigma_B$ is
the statistical uncertainty on the branching ratio due simply to the uncertainty in the signal yield.

\begin{figure} %fig 11
\centerline{\includegraphics[width=5.0in]{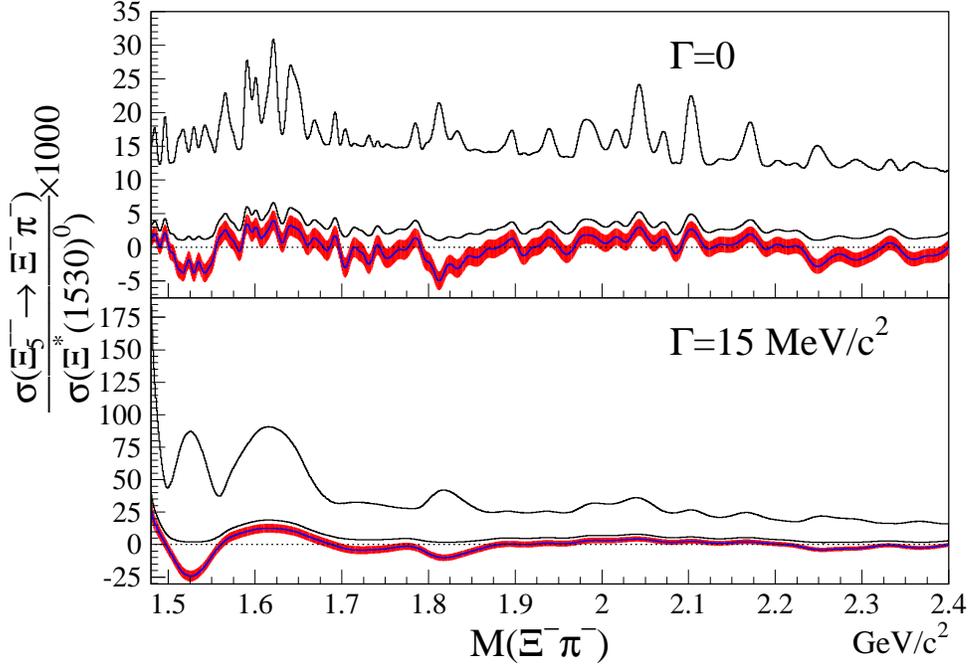}}
\caption{
$\frac{\sigma(\Xi_5^{--})\times \textrm{BR}(\Xi_5^{--}\to \Xi^-\pi^-)}{\sigma(\Xi^*(1530)^0)}$ versus mass.
Top (bottom) plots show results for a $\Xi_5^{--}$ natural width of $0$ ($15$~MeV/$c^2$).  
The shaded region encompasses the $1 \sigma$ statistical uncertainty with the central value in the middle.  
The top curve shows the 95\% CL upper limit including systematic uncertainties while 
the middle curve is the 95\% CL upper limit with statistical uncertainties only.}
\label{fig:xsecul_xistar}
\end{figure}

Figure~\ref{fig:xsecul_xistar} shows the results for 
$\frac{\sigma\left(\Xi_5^{--}\right) \cdot \textrm{BR}\left(\Xi_5^{--} \!\to\Xi^-\pi^-\right)}{\sigma\left(\Xi^*(1530)^0\right)}$
with an assumed natural width of 0 (15)~MeV/$c^2$ for the top (bottom) plot.  
This is the result corrected for all undetected particles.
The shaded band shows the $\pm1\sigma$ limits with statistical uncertainties only; 
the line in the middle of the band is the central value.  
The top curve shows the 95\% CL upper limit using the method described above including statistical and
systematic uncertainties.  
The curve between the full upper limit and the $1\sigma$ band is the 95\% CL upper limit using
the method described above with no systematic uncertainties included.  
The large systematic uncertainties are due to the attempt to correct for the significant fraction of particles 
outside of our acceptance.  
While this systematic uncertainty significantly degrades the limit, 
the production times branching ratio of the pentaquark relative to $\Xi^*(1530)^0$ production is still 
less than $0.032$ $(0.091)$ at 95\% CL over the 
%entire 
mass range 
1.5 to 2.4 \gevcc
for a natural width of 0 (15)~MeV/$c^2$.  
The background, as shown in Figure~\ref{fig:result:wrongsign_fit}, is rising so rapidly below 1.5 
\gevcc that fitting becomes very sensitive to the form assumed for the
background.  Consequently we do not quote upper limits below 1.5 \gevcc.

\begin{figure} %fig 12
\centerline{\includegraphics[width=5.0in]{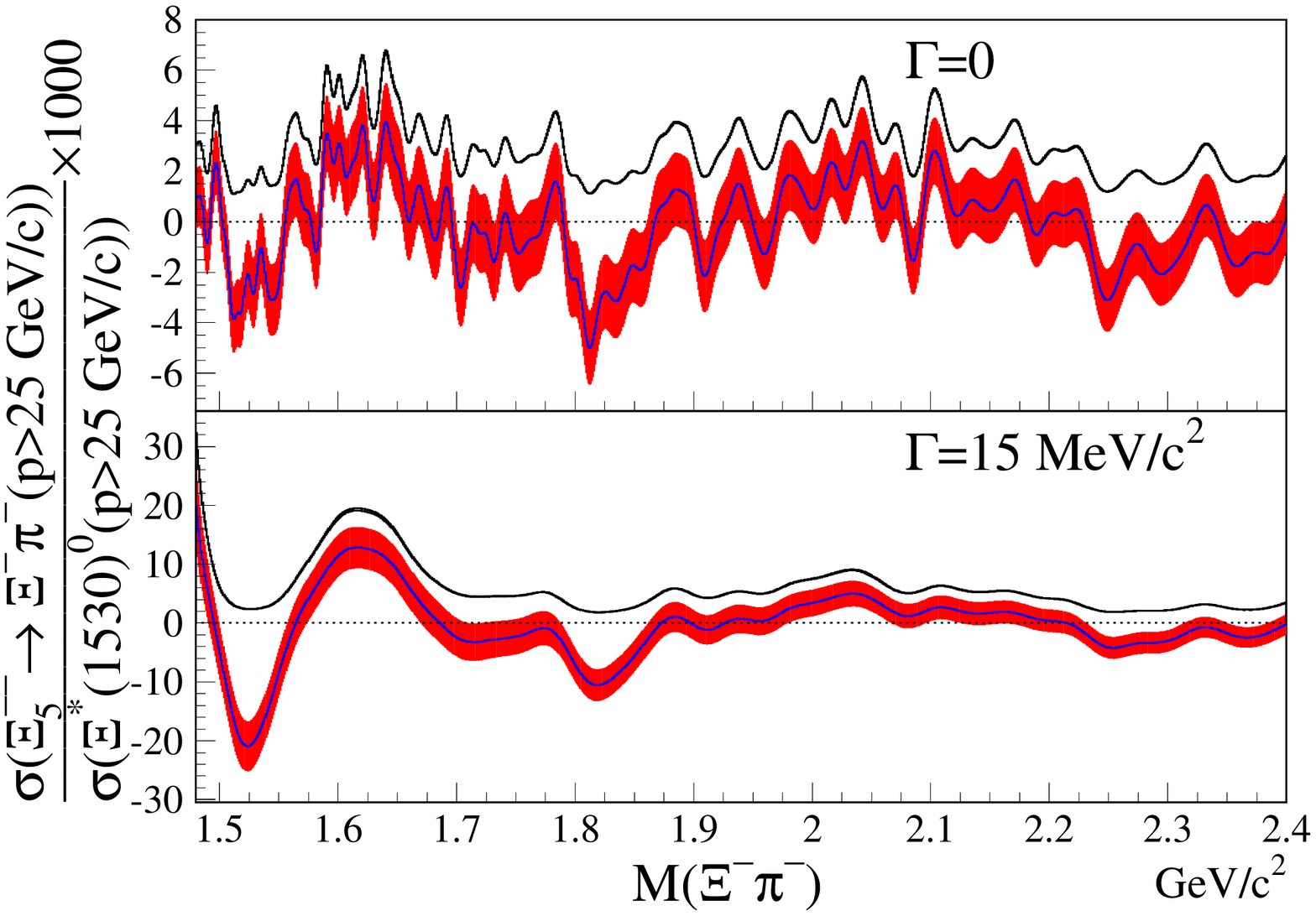}}
\caption{
$\frac{\sigma(\Xi_5^{--})\times \textrm{BR}(\Xi^{--}\to \Xi^-\pi^-)}{\sigma(\Xi^*(1530)^0)}$ 
for $p>25$~GeV/$c$ versus mass.
Top (bottom) plots show results for a $\Xi_5^{--}$ natural width of $0$ ($15$~MeV/$c^2$).  
The shaded region encompasses the $1 \sigma$ statistical uncertainty with the central value in the middle.
The top curve shows the 95\% CL upper limit including systematic uncertainties and is 
virtually indistinguishable from the middle curve which shows the 95\% CL upper limit with 
statistical uncertainties only.}
\label{fig:xsecula_xistar}
\end{figure}

The plots in Figure~\ref{fig:xsecula_xistar} show the same results for the restricted range of momentum greater
than 25 GeV/$c$.  That is, they show limits on relative cross sections for particles 
($\Xi_5^{--}$,  $\Xi^*(1530)^0$) produced with $p > 25$ \gevc.

\section{Conclusions}

We find no evidence for pentaquarks decaying to $\Xi^-\pi^-$ in the mass range of
1480~MeV/$c^2$ to 2400~MeV/$c^2$.  
In contrast, we observe 
about 65\,000 
$\Xi^*(1530)^0 \to \Xi^- \pi^+$ particles which have
a very similar topology and energy release.
We set 95\% CL upper limits on the yield over the entire mass range with a maximum of 600 (3000) 
events for an assumed natural width of 0 (15)~MeV/$c^2$.  
We also obtain 95\% CL upper limits on the cross section for pentaquark production times 
the branching ratio to $\Xi^-\pi^-$ relative to $\Xi^*(1530)^0 \to \Xi^- \pi^+$.  
These limits are determined for two cases.  
The first case is for parent particles produced at any momenta where we find a maximum upper limit of 
$\frac{\sigma\left(\Xi_5^{--}\right) \cdot \textrm{BR}\left(\Xi_5^{--} \!\to 
\Xi^-\pi^-\right)}{\sigma\left(\Xi^*(1530)^0\right)} < 0.032\; (0.091)$
at 95\% CL for a natural width of 0 (15)~MeV/$c^2$.
In the second case we measure the relative cross sections for parent particles with momenta above 25~GeV/$c$ 
(a region of good acceptance) and calculate 95\% CL limits of 
$\frac{\sigma\left(\Xi_5^{--}\right) \cdot \textrm{BR}\left(\Xi_5^{--} \!\to
\Xi^-\pi^-\right)}{\sigma\left(\Xi^*(1530)^0\right)} < 0.007\; (0.019)$
for a natural width of 0 (15)~MeV/$c^2$.  

The only experiment reporting an observation of the $\Xi_5^{--}$\ is NA49~\cite{Alt:2003vb} which shows about 15
$\Xi^*(1530)^0 \to \Xi^- \pi^+$ candidates, while reconstructing 38  $\Xi_5^{--}\to\Xi^- \pi^-$ candidates.
The FOCUS results for photon interactions
presented here represent 
%%%%$\Xi^*(1530)^0$ 
samples that are more than 4000 times larger, show no evidence
for a state $\Xi_5^{--} \!\to\Xi^-\pi^-$, and are 
%in clear disagreement with NA49.
in marked contrast with the NA49 results for pp interactions.

\section{Acknowledgments}
We wish to acknowledge the assistance of the staffs of Fermi National
Accelerator Laboratory, the INFN of Italy, and the physics departments
of the collaborating institutions. This research was supported in part
by the U.~S.  National Science Foundation, the U.~S. Department of
Energy, the Italian Istituto Nazionale di Fisica Nucleare and
Ministero dell'Istruzione dell'Universit\`a e della Ricerca, the
Brazilian Conselho Nacional de Desenvolvimento Cient\'{\i}fico e
Tecnol\'ogico, CONACyT-M\'exico, the Korean Ministry of Education, 
and the Korean Science and Engineering Foundation.

\bibliographystyle{h-elsevier_ed}
\bibliography{pentaquark}

\end{document}